\documentclass[a4paper,11pt]{article}
\pdfoutput=1 % if your are submitting a pdflatex (i.e. if you have
\usepackage{jheppub} % for details on the use of the package, please
\usepackage{graphicx}
\usepackage[T1]{fontenc} % if needed
\usepackage{amssymb,amsmath,bm,natbib}
\usepackage{color}
\usepackage{slashed}
\usepackage{graphics}
\usepackage{graphicx}
\usepackage[utf8]{inputenc}
\usepackage[caption=false]{subfig}
\usepackage{hyperref}
\usepackage{url}
\usepackage{dsfont}
\usepackage{float}
\usepackage{cancel}
\usepackage{units}
\usepackage{blindtext}
\usepackage[utf8]{inputenc}
\usepackage{upgreek}
\usepackage{booktabs}
\usepackage[dvipsnames,table,xcdraw]{xcolor}
\usepackage{enumerate}

\usepackage[normalem]{ulem}

\renewcommand{\eqref}[1]{\mbox{eq.~(\ref{#1})}}

\title{Modified-gravity theories with nondynamical background fields}

\author[a]{Carlos M. Reyes}
\author[b]{and Marco Schreck}

\affiliation[a]{Centro de Ciencias Exactas, Universidad del B\'{i}o-B\'{i}o, Avda. Andr\'es Bello 720, 3800708, Chill\'{a}n, Chile}
\affiliation[b]{Departamento de F\'{i}sica, Universidade Federal do Maranh\~{a}o, Campus Universit\'{a}rio do Bacanga, S\~{a}o Lu\'{i}s (MA), 65085-580, Brazil}

\emailAdd{creyes@ubiobio.cl}
\emailAdd{marco.schreck@ufma.br}

\abstract{We study the dynamics of a modified-gravity theory, which is supplemented by an extended Gibbons-Hawking-York boundary term and incorporates diffeomorphism
violation through nondynamical background fields denoted as $u$ and $s^{\mu\nu}$ in the literature. An ADM decomposition allows us to project the modified Einstein equations into purely spacelike hypersurfaces, which implies the field equations for the induced dynamical three-metric. We also obtain the Hamilton-Jacobi equations of motion for the canonical variables of the theory based on its Hamiltonian, which was derived in a previous work.
The computations show that the dynamical field equations obtained from the Lagrangian and Hamiltonian approaches are consistent with each other. Connections to Brans-Dicke theory and ghost-free massive gravity are established.}


\begin{document}
\maketitle
\flushbottom
%....................................................................................
\section{Introduction}
\label{sec:introduction}
%....................................................................................
The phenomena of perihelion advance of Mercury~\cite{Einstein:1915bz}, light deflection~\cite{Dyson:1920cwa}, and gravitational redshift~\cite{Pound:1960zz} compose the three classical tests of General Relativity (GR) and successfully confirm the latter as the correct gravity theory. Further significant validations of GR are the detection of the geodetic and frame-dragging effects by Gravity Probe B \cite{Everitt:2011hp}, the first direct observation of gravitational waves by LIGO \cite{LIGOScientific:2016aoc}, and the image of the accretion disk of the supermassive central black hole in the elliptic galaxy M87 recently taken by the Event Horizon Telescope~\cite{EventHorizonTelescope:2019dse}. Apart from linearized GR exhibiting a propagating massless spin-2 excitation, diffeomorphism invariance is at the heart of Einstein's theory. This symmetry encodes the profound property of GR being invariant under nonlinear, differentiable maps from a spacetime manifold onto itself. The field equations of GR are also covariant under general coordinate transformations, but the latter only operate on the atlas of a manifold as opposed to points, sets, and curves on the manifold proper. The geometry of a curved spacetime manifold is locally equivalent to the pseudo-Euclidean geometry of Minkowski spacetime, whose symmetry group is the Poincar\'{e} group. Our focus is on diffeomorphisms instead of the local symmetry structure of GR.

Any departure from diffeomorphism invariance in a modified-gravity theory is expected to have significant theoretical and observational implications. In particular, the constraint algebra \cite{Gambini:1996} and Dirac observables \cite{Giesel:2008zz} of GR are likely to be modified. The exact physical consequences of such deviations are challenging to identify and extremely difficult to measure directly, as manipulations of the spacetime manifold are beyond the capabilities of presently conceivable experiments. However, parameterizing explicit diffeomorphism violation via nondynamical background fields, which have nonstandard transformation properties, can provide a window into probing associated phenomena.

In principle, diffeomorphism violation is as old as GR itself. While his theory was still in the making, Einstein required that the determinant of the spacetime metric correspond to unity, which led to vast simplifications of his field equations~\cite{Einstein:1916}.
This condition is also incorporated in unimodular gravity~\cite{Buchmuller:1988wx,Unruh:1988in,Bufalo:2015wda}, which is not invariant under the full diffeomorphism group of GR, but only under volume-preserving diffeomorphisms. Unimodular gravity corresponds to GR with an arbitrary cosmological constant, although the latter arises as an integration constant instead of a parameter introduced at the level of the action~\cite{Ng:1990xz,Finkelstein:2000pg}.

Despite the beauty of Einstein's gravity theory and its astounding achievements, alterations are expected when the latter is to be reconciled with quantum theory. Modifications of the gravitational laws of physics have been found to occur naturally at the Planck scale, where quantum effects are expected to become dominant. Such settings are provided by string theory, branes, and supergravity. For example, diffeomorphism violation in gravity can arise from a spontaneous breaking of Lorentz invariance in particular string field theories~\cite{Kostelecky-Samuel}. The potential mechanism gives rise to problems related to compactified dimensions in the context of inflation, suggesting the generation of mass terms for some metric components~\cite{Phen1,Phen2}. Recently, other diffeomorphism-violating gravity models were shown to involve such mass terms~\cite{Review1,Review2}.

Generically, both spontaneous and explicit diffeomorphism symmetry breaking are parameterized within the comprehensive effective-field theory framework known as the gravitational Standard-Model Extension (SME) \cite{Kostelecky:2003fs,Kostelecky:2020hbb}. Searches for diffeomorphism violation --- let it be spontaneous or explicit --- can constrain SME coefficients. The direct detection of gravitational waves by LIGO \cite{LIGOScientific:2016aoc} ushered in the new era of gravitational-wave astronomy. The latter constitutes one method to probe deviations from GR, which has gained vast prominence and has been providing an ever-increasing number of SME constraints~\cite{Kostelecky:2016kfm,Mewes:2019dhj,Shao:2020shv,Wang:2021ctl,ONeal-Ault:2021uwu,Niu:2022yhr}. The absence of energy losses of particles by gravitational Cherenkov radiation \cite{Moore:2001bv,Kostelecky:2015dpa} provides an alternative possibility of bounding SME coefficients. Some experimental constraints on background field coefficients such as those obtained in, e.g., refs.~\cite{Kostelecky:2021tdf,Ivanov:2021bvk} are interpreted as bounds on explicit diffeomorphism violation. The data tables \cite{Kostelecky:2008bfz} can be consulted as an extensive compilation of such constraints.

Diffeomorphism invariance is also violated explicitly in particular models of massive gravity. An extension of GR implementing the propagation of a massive spin-2 excitation is known as massive gravity and its construction has received significant attention during the past years. The action initially introduced by Fierz and Pauli \cite{FP} is ghost-free at the linearized level, but does not agree with GR in the massless limit. The Vainshtein mechanism~\cite{Vainshtein:1972sx} solves this problem by taking into account nonlinear effects. Unfortunately, with that settled, another severe issue was revealed by Boulware and Deser: the occurrence of a ghost mode at the nonlinear level at an unacceptably small scale \cite{BD}.

Around ten years ago, the construction of de Rham, Gabadadze, and Tolley (dRGT)~\cite{deRham:2010kj} led to a breakthrough, since it avoids the Boulware-Deser (BD) ghost in the decoupling limit to all orders. This realm of research has uncovered many interesting aspects such as its interplay with the physics of diffeomorphism violation~\cite{Hassan,Mattingly,Kostelecky:2021xhb}, extra dimensions~\cite{DGP1,DGP2}, the St\"uckelberg trick~\cite{deRham:2011rn,StuckHassan,Siegel,ADMStuck}, linearized~\cite{FP,Bergshoeff:2012ud} vs.~nonlinear massive gravity~\cite{BD,Vainshtein:1972sx,deRham:2010kj,Nonlinear1,Nonlinear2,deRham:2011ca,Hassan:2011vm}, the helicity decomposition of the metric perturbation \cite{deRham:2011qq}, and the Ostrogradsky formalism as well as ghost states~\cite{Stelle,BD,Hassan:2011ea}. It is worthwhile to consult the valuable review \cite{deRham:2014zqa} for this research area.

The motivations for constructing gravity theories beyond GR are multifold. Low-energy extensions of Einstein's gravity are basically motivated by cosmological observations. Understanding the late-time acceleration of the Universe and inflation as well as the search for dark matter has inspired various modifications of GR~\cite{Buchdahl:1970ynr,Starobinsky:1980te,f(R),DeFelice:2010aj,Horn,Galileons}. Some alternatives such as Brans-Dicke theory \cite{Brans:1961sx} intend to incorporate Mach's principle into the description of gravity. The latter was one of Einstein's motivations for developing his GR, but it seems that he abandoned this goal later on. Ho\v{r}ava-Lifshitz gravity \cite{Horava:2008ih,Horava:2009uw} is formulated to improve the UV behavior of the graviton propagator, while the theory approaches Einstein's gravity in the IR regime. Another interesting model is cardinal gravity \cite{Kostelecky:2009zr,Seifert:2019kuz} that interprets long-range gravitational forces as propagating Nambu-Goldstone modes arising from spontaneous symmetry breaking. In general, some extensions of GR introduce operators with higher-order derivatives \cite{LW-cosmology,DHOST1,DHOST2}, others involve functions of the Ricci curvature scalar~\cite{Buchdahl:1970ynr,Starobinsky:1980te,f(R),DeFelice:2010aj} or additional vector and tensor fields coupling to spacetime curvature \cite{Gripaios:2004ms,Bluhm:2004ep,Bluhm:2007bd,Seifert:2009vr,Colladay:2019lig,Seifert:2019xan,Bonder:2021gjo,Delhom:2022}. Stating an exhaustive list of references on extensions or modifications of Einstein's relativity lies beyond our scope. References~\cite{Will:2014kxa,Heisenberg:2018vsk,Tasson:2016xib} provide elaborate reviews on modified-gravity theories and experimental tests where ref.~\cite{Tasson:2016xib} focuses on the SME, in particular.

In light of the previously mentioned lines of research, many modified-gravity theories exhibit diffeomorphism violation. While spontaneous symmetry violation such as in Refs.~\cite{Gripaios:2004ms,Bluhm:2004ep,Bluhm:2007bd,Seifert:2009vr,Colladay:2019lig,Seifert:2019xan,Bonder:2021gjo,Delhom:2022} is known to be benign \cite{Bluhm:2014oua}, the question may arise whether or not explicit diffeomorphism violation implies internal inconsistencies and under which circumstances. This paper intends to answer that question for a particular class of modified-gravity theories. Its action involves coordinate scalars beyond GR being noninvariant under diffeomorphisms. Symmetry violation is parameterized by a nondynamical scalar background field denoted as $u$ and a tensor-valued one called $s^{\mu\nu}$, respectively. The latter settings and extensions thereof are part of the gravitational SME \cite{Kostelecky:2003fs,Kostelecky:2020hbb} mentioned before and have been subject to phenomenological \cite{Bailey:2006fd,Bailey:2009me,Tso:2011up,Bailey:2013oda,Bonder:2020fpn} as well as theoretical studies \cite{Bonder:2015maa}. As these alterations of GR are quite generic, relationships to specific modified-gravity theories in the literature are expected. In particular, we will refer to Brans-Dicke theory and dRGT massive gravity.

In the recent work~\cite{Reyes:2021cpx}, the formalism developed by Arnowitt, Deser, and Misner (ADM)~\cite{Arnowitt:1962hi,Misner:1973,Hanson:1976,Poisson:2002,Poisson:2004,Gourgoulhon:2007ue,Arnowitt:2008,Gourgoulhon:2012} was applied to the class of modified-gravity theories characterized by $u$ and $s^{\mu\nu}$. The calculations showed that the background fields imply an altered constraint structure in the Hamiltonian formulation.
Incorporating an extension of the Gibbons-Hawking-York (GHY) boundary term \cite{York:1972sj,Gibbons:1976ue} to eliminate additional second-order time derivatives of the metric, which occur in the Lagrangian and couple to the background fields, was an essential ingredient for constructing the Hamiltonian. Our objective in the present article is to understand the dynamics of this theory better. We will be working both within the covariant and the Hamiltonian formulation. We also note in passing that the authors of ref.~\cite{ONeal-Ault:2020ebv} investigate a similar, though not equivalent, theory. Their interest and approach differ from ours in certain aspects.

Our paper is organized as follows. In section~\ref{sec:modified-gravity} we recapitulate the modified-gravity theory that forms the foundation of ref.~\cite{Reyes:2021cpx} and define the basic variables employed in the ADM decomposition. Furthermore, we emphasize correspondences between the latter setting and Brans-Dicke theory and dRGT massive gravity. In section~\ref{sec:momenta-hamiltonians}, the canonical-momentum variables and the Hamiltonians are compiled from ref.~\cite{Reyes:2021cpx}. Our new investigations start in section~\ref{eq:projected-field-equations} where the modified Einstein equations are projected into purely spacelike hypersurfaces of the spacetime foliation. Section~\ref{sec:hamilton-jacobi} is dedicated to the Hamilton-Jacobi equations of motion. Our essential finding here is that the latter correspond to the projected field equations in the covariant formulation. Finally, via section~\ref{sec:outlook} we would like to motivate further SME-related studies along the same line by demonstrating which SME background fields play a role in linearized dRGT theory and Ho\v{r}ava-Lifshitz gravity. Our findings are concluded on in section~\ref{sec:conclusions}. We employ natural units with $c=1$ unless otherwise stated. Our metric signature is $(-,+,+,+)$. Greek indices are spacetime indices, whereas Latin indices describe quantities living in purely spacelike hypersurfaces of the ADM decomposition. The \textit{Mathematica} package \textit{xTensor} \cite{xTensor:2020} provides computational support.

%................................................................................
\section{Modified gravity with background fields}
\label{sec:modified-gravity}
%................................................................................
We focus on the following modified Einstein-Hilbert (EH) action \cite{Kostelecky:2003fs,Kostelecky:2020hbb} without a cosmological constant:
\begin{equation}
\label{eq:minimal-gravity-sme-reformulated}
S=\int_{\mathcal{M}}\mathrm{d}^4x\frac{\sqrt{-g}}{2\kappa}\left[(1-u){}^{(4)}R+s^{\mu\nu}{}^{(4)}R_{\mu\nu}\right]\,,
\end{equation}
with $\kappa=8\pi G_N$, the Ricci tensor ${}^{(4)}R_{\mu\nu}$ and the associated Ricci scalar ${}^{(4)}R:={}^{(4)}R^{\mu}_{\phantom{\mu}\mu}$ of the four-dimensional spacetime manifold $\mathcal{M}$ with metric tensor $g_{\mu\nu}$ and $g:=\det(g_{\mu\nu})$. Furthermore, $u=u(x)$ and $s^{\mu\nu}=s^{\mu\nu}(x)$ are nondynamical background fields having a generic spacetime dependence. The latter are considered to have zero fluctuations: $\delta u=0$ and $\delta s^{\mu\nu}=0$. All fields are defined in the tangent bundle of $\mathcal{M}$. The modified Einstein equations for \eqref{eq:minimal-gravity-sme-reformulated} were first obtained in ref.~\cite{Bailey:2006fd} and can be read from eqs.~(6), (7) of the latter reference:
\begin{align}
\label{eq:einstein-equations-modified-generic}
0&=(1-u){}^{(4)}G^{\mu\nu}+\frac{1}{2}(\nabla^{\mu}\nabla^{\nu}u+\nabla^{\nu}\nabla^{\mu}u)-g^{\mu\nu}\square u \notag \\
&\phantom{{}={}}-\frac{1}{2}\left(s^{\alpha\beta}{}^{(4)}R_{\alpha\beta}g^{\mu\nu}+\nabla_{\alpha}\nabla^{\mu}s^{\alpha\nu}+\nabla_{\alpha}\nabla^{\nu}s^{\alpha\mu}-\square s^{\mu\nu}-g^{\mu\nu}\nabla_{\alpha}\nabla_{\beta}s^{\alpha\beta}\right)\,,
\end{align}
where ${}^{(4)}G^{\mu\nu}$ is the Einstein tensor. Furthermore, $\nabla_{\mu}$ is the covariant derivative on the spacetime manifold $\mathcal{M}$ and $\square:=\nabla_{\mu}\nabla^{\mu}$ the d'Alembertian. The Bianchi identities of Riemannian geometry imply nontrivial restrictions on the form of the background fields $u$ and $s^{\mu\nu}$, which is a well-known feature of this theory that has been indicated in a series of papers such as refs.~\cite{Kostelecky:2003fs,Kostelecky:2020hbb,Bluhm:2014oua}, amongst others. We intend to study the consequences of these restrictions in a future paper, but our current focus is on the purely dynamical aspects of \eqref{eq:minimal-gravity-sme-reformulated}.

We use the ADM formalism expressed in terms of the configuration space variables $X:=\{N,N^i,q_{ij}\}$. Here, $N$ is the lapse function, $N^i$ are the components of the shift vector, and $q_{ij}$ are the components of the induced metric on a spacelike hypersurface $\Sigma_t$~\cite{Arnowitt:1962hi,Misner:1973,Hanson:1976,Poisson:2002,Poisson:2004,Gourgoulhon:2007ue,Arnowitt:2008,Gourgoulhon:2012}. It then holds that $\sqrt{-g}=N\sqrt{q}$ with $q:=\det(q_{ij})$ and
\begin{equation}
g_{00}=-N^2+q^{ij}N_iN_j\,,\quad g_{0i}=N_i\,,\quad g_{ij}=q_{ij}\,.
\end{equation}
Furthermore, the components of the contravariant metric amount to
\begin{equation}
g^{00}=-\frac{1}{N^2}\,,\quad g^{0i}=\frac{N^i}{N^2}\,,\quad g^{ij}=q^{ij}-\frac{N^iN^j}{N^2}\,.
\end{equation}
Since we will have to study projections of the modified Einstein equations into $\Sigma_t$, recall also the following decomposition formula for the Ricci tensor and scalar curvature that arise from the Gauss-Codazzi equation:
\begin{subequations}
\label{eq:decompositions-ricci}
\begin{align}
q^i_{\phantom{i}\alpha}q^j_{\phantom{j}\beta}{}^{(4)}R^{\alpha\beta}&=R^{ij}+\frac{1}{N}q^{ia}q^{jb}\mathcal{L}_mK_{ab}-\frac{1}{N}D^iD^jN+KK^{ij}-2K^{ik}K_k^{\phantom{k}j} \notag \\
&=\frac{1}{N}q^{ia}q^{jb}\mathcal{L}_mK_{ab}-(D^ia^j+a^ia^j)+R^{ij}+KK^{ij}-2K^{ik}K_k^{\phantom{k}j}\,, \\[2ex]
{}^{(4)}R&=R+K^2+K_{ij}K^{ij}-\frac{2}{N}D_iD^iN+\frac{2}{N}\mathcal{L}_mK\,,
\end{align}
\end{subequations}
in terms of the extrinsic curvature $K_{ab}:=D_an_b$ with $n_{\mu}=(-N,0,0,0)$, the Ricci tensor $R_{ij}$, and the Ricci scalar $R$ in a spacelike hypersurface $\Sigma_t$. Furthermore, $\mathcal{L}_m$ denotes the Lie derivative \cite{Carroll:1997ar} along the four-vector $m^{\mu}:=Nn^{\mu}$ with $n^{\mu}=(1/N,-N^i/N)$ and $a_i:=D_i\ln(N)$ is known as the acceleration. Here, $D_i$ is the covariant derivative in $\Sigma_t$, which is compatible with the induced metric $q_{ij}$.

The background field $s^{\mu\nu}$ is tensor-valued and decomposes into three distinct sectors according to the foliation of the spacetime manifold in terms of the hypersurfaces $\Sigma_t$. The following identity is valuable for performing this decomposition:
\begin{equation}
\label{eq:decomposition-s}
s^{\alpha\beta}=q^{\alpha}_{\phantom{\alpha}\mu}q^{\beta}_{\phantom{\beta}\nu}s^{\mu\nu}-(q^{\alpha}_{\phantom{\alpha}\nu}n^{\beta}+q^{\beta}_{\phantom{\beta}\nu}n^{\alpha})s^{\nu \mathbf{n}}+n^{\alpha}n^{\beta}s^{\mathbf{nn}}\,,
\end{equation}
where $q^{\mu}_{\phantom{\mu}\nu}$ projects a tensor or a part of it into $\Sigma_t$. Here we define $s^{ij}:=q^i_{\phantom{i}\mu}q^j_{\phantom{j}\nu}s^{\mu\nu}$ as the purely spacelike sector of $s^{\mu\nu}$ that lives in $\Sigma_t$ entirely. Furthermore, let $s^{i\mathbf{n}}:=q^i_{\phantom{i}\mu}n_{\nu}s^{\mu\nu}$ be the vector-valued piece and $s^{\mathbf{n}\mathbf{n}}:=n_{\mu}n_{\nu}s^{\mu\nu}$ the scalar part.

Now, the total ADM-decomposed action based on \eqref{eq:minimal-gravity-sme-reformulated} reads \cite{Reyes:2021cpx}
\begin{subequations}
\label{eq:theory-definition}
\begin{equation}
S=\int_{\mathcal M}\mathrm{d}t\mathrm{d}^3x\,\mathcal{L}_{\mathrm{ADM}}\,,\quad \mathcal{L}_{\mathrm{ADM}}=\mathcal{L}^{(0)}+\mathcal{L}^{(u)}+
\sum_{i=1,2,3}\mathcal{L}^{(s)}_i\,,
\end{equation}
where $\mathcal{L}^{(0)}$ is the ADM-decomposed EH Lagrange density,
\begin{equation}
\label{eq:theory-einstein-hilbert}
\mathcal{L}^{(0)}=\frac{N\sqrt{q}}{2\kappa}\left(\frac{2}{N}\mathcal{L}_mK-\frac{2}{N}D_iD^iN+R+K^2+K_{ij}K^{ij}\right)\,,
\end{equation}
and the modifications are
\begin{align} \label{eq:theory-definition-u}
\mathcal{L}^{(u)}&=\frac{N\sqrt{q}}{2\kappa}\,\left[-u(R-K^2+K_{ij}K^{ij})+\frac{2}{N}(K\mathcal{L}_mu+uD_iD^iN)\right]\,, \displaybreak[0]\\[2ex]
\label{eq:theory-definition-snn}
\mathcal{L}^{(s)}_1&=\frac{N\sqrt{q}}{2\kappa}\left[-\frac{1}{N}(K_{ij}\mathcal{L}_ms^{ij}+s^{ij}D_iD_jN)+s^{ij}(R_{ij}-2K_i^{\phantom{i}l}K_{lj})\right]\,, \displaybreak[0]\\[2ex]
\label{eq:theory-definition-sij}
\mathcal{L}^{(s)}_2&=\frac{N\sqrt{q}}{2\kappa}\left[s^{\mathbf{nn}}\left(\frac{1}{N}D_iD^iN-K^{ij}K_{ij}+K^2\right)+\frac{1}{N}K\mathcal{L}_ms^{\mathbf{nn}}\right]\,, \displaybreak[0]\\[2ex]
\mathcal{L}^{(s)}_3&=\frac{N\sqrt{q}}{2\kappa}\left[2s^{i\mathbf{n}}(D_iK-D_lK^l_{\phantom{l}i})\right]\,.
\end{align}
\end{subequations}
By defining new shift vector components as $\tilde{N}^i:= N^i-Ns^{i\mathbf{n}}$, the Lagrange density $\mathcal{L}_3^{(s)}$, which involves the vector-valued coefficients $s^{i\mathbf{n}}$ only, can be reproduced at first order in $s^{i\mathbf{n}}$ from the ADM decomposition of the EH Lagrange density. This reason is a strong argument for $s^{i\mathbf{n}}$ corresponding to mere gauge degrees of freedom \cite{Reyes:2021cpx}. Thus, they will be completely discarded in the remainder of the paper. Then, we also take into account an extended GHY boundary term \cite{Reyes:2021cpx} of the form
\begin{equation}
\label{eq:modified-GHY}
S_{\substack{\text{ext} \\ \text{GHY}}}=\frac{\varepsilon}{2\kappa}\oint_{\partial\mathcal{M}} \mathrm{d}^3y\,\sqrt{q}\,\left[2(1-u)K-s^{\mathbf{nn}}K+K_{ij}s^{ij}\right]\,,
\end{equation}
where the parameter $\varepsilon=\mp 1$ for a spacelike (timelike) boundary $\partial\mathcal{M}$ of the spacetime manifold $\mathcal{M}$ and the integral runs over the coordinates $y^a$ defined on this boundary. Last but not least, we introduce a second boundary term for $u$ and $s^{\mathbf{nn}}$ that is of plainly different nature compared to that of~\eqref{eq:modified-GHY}:
\begin{equation}
\label{eq:additional-boundary-term}
S_{\partial\Sigma}=-\frac{1}{2\kappa}\oint_{\partial\Sigma_t} \mathrm{d}^2z\,\sqrt{q}r_l\left[ND^l(2u+s^{\mathbf{nn}})\right]\,,
\end{equation}
with the coordinates $z^a$ given on the boundary of a spacelike hypersurface $\Sigma_t$ and a suitably normalized vector $r_l$ orthogonal to the boundary. In earnest, such boundary terms can emerge when integrations by parts are performed over $\Sigma_t$ while computing the Hamiltonian from the corresponding Lagrangian via a Legendre transformation~\cite{Reyes:2021cpx}. The purpose of these boundary contributions will become clear later on.

Any modified-gravity theory that is formulated at the level of an action is expected to have a corresponding background field in the comprehensive parameterization of the gravitational SME recently developed in ref.~\cite{Kostelecky:2020hbb}. However, it may be challenging to establish such a correspondence for a specific model under consideration. In the following, a number of popular examples are provided, which shall serve as a motivation for developing a better understanding of nondynamical background fields in the gravitational SME.

\subsection{Brans-Dicke and $f(R)$ theories}

An early motivation for modifying Einstein's GR was to incorporate Mach's principle. One among many possibilities of formulating the latter is via a gravitational ``constant'' that is promoted to a dynamical field with an explicit dependence on the spacetime coordinates~\cite{Bondi:1996md}. In the formulation by Brans and Dicke~\cite{Brans:1961sx} the inverse of the gravitational constant is replaced by a scalar field $\phi$, whereupon the modified field equations are based on the variation of a modified action as follows:
\begin{equation}
0=\delta\int_{\mathcal{M}}\mathrm{d}^4x\,\sqrt{-g}\left(\phi R+16\pi L_m-\omega\frac{(\nabla\phi)^2}{\phi}\right)\,.
\end{equation}
Here, $\omega$ is a dimensionless free parameter and $L_m$ contains the matter fields. The action can then be cast into the form
\begin{equation}
\label{eq:action-brans-dicke}
S_{\mathrm{BD}}=\int_{\mathcal{M}}\mathrm{d}^4x\,\sqrt{-g}\left(\frac{G_N}{2\kappa}\phi R+L_m-\frac{\omega G_N}{2\kappa}\frac{(\nabla\phi)^2}{\phi}\right)\,,
\end{equation}
where $[G_N]=[\kappa]=-2$ and $[\phi]=[R]=2$. We identify
\begin{equation}
\label{eq:connection-phi-u}
G_N\phi=1-u\,,
\end{equation}
with the scalar background field $u$ in $\mathcal{L}^{(u)}$ of \eqref{eq:theory-definition-u}. A reformulation of the dynamical term in \eqref{eq:action-brans-dicke} amounts to
\begin{equation}
S_{\mathrm{BD}}\supset \int_{\mathcal{M}}\mathrm{d}^4x\,\sqrt{-g}\left(-\frac{\omega G_N}{2\kappa}\frac{(\nabla\phi)^2}{\phi}\right)=-\int_{\mathcal{M}}\mathrm{d}^4x\,\frac{\sqrt{-g}}{2\kappa}\frac{(\nabla u)^2}{1-u}\,.
\end{equation}
Now, the modified field equations \cite{Brans:1961sx} following from \eqref{eq:action-brans-dicke} read
\begin{align}
G_N\phi G_{\mu\nu}&=8\pi G_N T_{\mu\nu}+\frac{\omega}{G_N\phi}\left[\nabla_{\mu}(G_N\phi)\nabla_{\nu}(G_N\phi)-\frac{1}{2}g_{\mu\nu}(\nabla G_N\phi)^2\right] \notag \\
&\phantom{{}={}}+\nabla_{\mu}\nabla_{\nu}G_N\phi-g_{\mu\nu}\square G_N\phi\,,
\end{align}
and expressing $\phi$ in terms of $u$ according to \eqref{eq:connection-phi-u} implies
\begin{align}
(1-u)G_{\mu\nu}&=8\pi G_N T_{\mu\nu}+\frac{\omega}{1-u}\left[\nabla_{\mu}u\nabla_{\nu}u-\frac{1}{2}g_{\mu\nu}(\nabla u)^2\right]-\nabla_{\mu}\nabla_{\nu}u+g_{\mu\nu}\square u\,.
\end{align}
Dropping the energy-momentum tensor as well as the dynamical contributions proportional to $\omega$ leads to the modified Einstein equations for the $u$ term. They are deduced from \eqref{eq:einstein-equations-modified-generic} by setting $s^{\mu\nu}=0$ and will be recalled in \eqref{eq:modified-einstein-u} to come. Thus, Brans-Dicke theory without the dynamical term can be identified with the nondynamical $u$ sector of the gravitational SME. To the best of our knowledge, this correspondence has not been stated in the literature, so far.

Another observation is that the Lagrange density $\mathcal{L}^{(u)}$ of \eqref{eq:theory-definition-u} corresponds, in principle, to a particular $f(R)$ theory involving a nondynamical background field such that $f(R)=(1-u)R$. It is not difficult to check that the field equations quoted in \eqref{eq:modified-einstein-u} are equivalent to eqs.~(1.1), (A4) of ref.~\cite{Buchdahl:1970ynr} with the energy-momentum tensor $T_{\mu\nu}=0$.

\subsection{dRGT massive gravity}

As GR, massive-gravity theories constructed before the seminal work \cite{deRham:2010kj} did not involve canonical momenta associated with the lapse function and the shift vector, which gives rise to 4 primary first-class constraints. Thus, counting the number of physical degrees of freedom \cite{Henneaux:1992} amounts to 6, which is one more than expected for a massive graviton. The additional degree of freedom corresponds to the BD ghost mode propagating at energy scales proportional to the minuscule graviton mass, which is unacceptable. dRGT theory emerged in an attempt to conceive a massive-gravity theory that is devoid of the BD ghost. To do so, the Einstein-Hilbert term is supplemented by a particular potential of the graviton that has a highly peculiar form. The theory is constructed in a manner to provide another primary and an additional secondary constraint \cite{Hassan:2011ea} on the three-metric components. Thereupon, a counting of the physical degrees of freedom indicates that the ghost mode is eliminated.

The potential of dRGT theory involves traces of coordinate two-tensors $\mathds{X}^{\mu}_{\phantom{\mu}\nu}:=\sqrt{g^{\mu\alpha}f_{\alpha\nu}}$ expressed in terms of the dynamical spacetime metric $g_{\mu\nu}$ and a nondynamical reference metric, which is frequently denoted as $f_{\mu\nu}$ \cite{deRham:2014zqa}. At the level of the pure-gravity action of dRGT theory it is challenging to identify the nondynamical background $f_{\mu\nu}$ or contractions thereof with SME coefficients. After all, the potential is not a simple sum of contributions involving $f_{\mu\nu}$ contracted with dynamical fields or curvature components in a linear fashion. The two-tensor $\mathds{X}^{\mu}_{\phantom{\mu}\nu}$ cannot be evaluated generically, but must be computed based on particular choices of the dynamical metric and the background field.

Finding an appropriate map in the matter sector is plainly simpler. Coupling a particular set of matter fields to the gravity sector via both metrics $g_{\mu\nu}$ and $f_{\mu\nu}$ was demonstrated to reintroduce the BD ghost~\cite{deRham:2014naa}. A reemergence of the ghost can only be avoided by coupling in a more sophisticated way via an effective metric given by
\begin{equation}
\label{eq:effective-metric-dRGT}
g_{\mu\nu}^{(\mathrm{eff})}=\alpha^2g_{\mu\nu}+2\alpha\beta g_{\mu\alpha}\mathds{X}^{\alpha}_{\phantom{\alpha}\nu}+\beta^2f_{\mu\nu}\,,
\end{equation}
with arbitrary dimensionless parameters $\alpha,\beta$ that may be tuned to either couple with the dynamical metric or the reference metric or both. Now, a background vierbein $\overline{v}_{\mu}^{\phantom{\mu}a}$ is introduced such that $f_{\mu\nu}=\overline{v}_{\mu}^{\phantom{\mu}a}\overline{v}_{\nu}^{\phantom{\nu}b}\eta_{ab}$ with the Minkowski metric $\eta_{ab}$ in a local inertial frame. If the dynamical vierbein $e_{\mu}^{\phantom{\mu}a}$ and the background vierbein satisfy $e^{\mu}_{\phantom{\mu}a}\overline{v}_{\mu b}=e^{\mu}_{\phantom{\mu}b}\overline{v}_{\mu a}$, it holds that $\mathds{X}^{\mu}_{\phantom{\mu}\nu}=e^{\mu}_{\phantom{\mu}a}\overline{v}_{\nu}^{\phantom{\nu}a}=:\gamma^{\mu}_{\phantom{\mu}\nu}$ \cite{Hinterbichler:2012cn}. In the context of the gravitational SME, a redefinition of the (inverse) metric according to $g^{\mu\nu}\simeq (1+u)\tilde{g}^{\mu\nu}+s^{\mu\nu}$ with a new metric $\tilde{g}_{\mu\nu}$ can be carried out to remove $u$ and $s^{\mu\nu}$ from the pure-gravity sector. Expressing the former metric $g_{\mu\nu}$ in terms of the new metric $\tilde{g}_{\mu\nu}$ in the matter sector, introduces appropriate SME coefficients in the matter sector. By taking $s^{\mu\nu}$ as traceless, for the effective metric of \eqref{eq:effective-metric-dRGT} it is possible to identify \cite{Bluhm:2019ato}
\begin{equation}
u=-\frac{1}{2}\beta\gamma^{\sigma}_{\phantom{\sigma}\sigma}\,,\quad s^{\mu\nu}=-2\beta\left(\gamma^{\mu\nu}-\frac{1}{4}\gamma^{\sigma}_{\phantom{\sigma}\sigma}g^{\mu\nu}\right)\,.
\end{equation}
Thus, the theory of \eqref{eq:theory-definition} has a direct link to matter-gravity couplings in dRGT massive gravity.

\section{Canonical momenta and Hamiltonians}
\label{sec:momenta-hamiltonians}

We are now interested in understanding the dynamics of the modified-gravity theory based on \eqref{eq:minimal-gravity-sme-reformulated}. Both the dynamical equations of motion and the constraints are encoded in the modified Einstein equations. In principle, each constraint arises from \eqref{eq:einstein-equations-modified-generic} by applying suitable normal and tangential projections with respect to the hypersurfaces $\Sigma_t$ defined in the ADM formalism.

We now consider the set of momenta $\Pi:=\{\pi_N,\pi_i,\pi^{ij}\}$ associated with the canonical variables. They are defined as
\begin{equation}
\pi_N:=\frac{\partial\mathcal{L}_{\mathrm{ADM}}}{\partial \dot{N}}\,,\quad \pi_i:=\frac{\partial\mathcal{L}_{\mathrm{ADM}}}{\partial \dot{N}^i}\,,\quad \pi^{ij}:=\frac{\partial\mathcal{L}_{\mathrm{ADM}}}{\partial\dot{q}_{ij}}\,,
\end{equation}
where a dot stands for the time derivative. Since the Lagrangian of \eqref{eq:theory-definition} does not involve velocities of the gauge variables, $\dot{N}$ and $\dot{N}^i$, the canonical momentum conjugates to the lapse function and the shift vector, respectively, are found to be weakly equal to zero:
\begin{equation}
\pi_N\approx 0\,,\quad \pi_i\approx 0\,.
\end{equation}
Furthermore, we impose the following fundamental equal-time Poisson brackets between the canonically conjugate variables (see ref.~\cite{Bertschinger:2002}):
\begin{subequations}
\begin{align}
\{N(\mathbf{x}),\pi_N(\mathbf{x}')\}&=\delta^{(3)}(\mathbf{x}-\mathbf{x}')\,, \displaybreak[0]\\[2ex]
\{N^j(\mathbf{x}),\pi_i(\mathbf{x}')\}&=\delta_i^{\phantom{i}j}\delta^{(3)}(\mathbf{x}-\mathbf{x}')\,, \displaybreak[0]\\[2ex]
\{q_{kl}(\mathbf{x}),\pi^{ij}(\mathbf{x}')\}&=(\delta^i_{\phantom{i}k}\delta^j_{\phantom{j}l}-\delta^i_{\phantom{i}l}\delta^j_{\phantom{j}k})\delta^{(3)}(\mathbf{x}-\mathbf{x}')\,,
\end{align}
\end{subequations}
where the Poisson bracket is defined as
\begin{equation}
\{Q(\mathbf{x}),P(\mathbf{x}')\}:=\int_{\Sigma_t}\mathrm{d}^3\tilde{x}\,\left[\frac{\delta Q(\mathbf{x})}{\delta X_i(\tilde{\mathbf{x}},t)}\frac{\delta P(\mathbf{x}')}{\delta\Pi^i(\tilde{\mathbf{x}},t)}-\frac{\delta Q(\mathbf{x})}{\delta\Pi^i(\tilde{\mathbf{x}},t)}\frac{\delta P(\mathbf{x}')}{\delta X_i(\tilde{\mathbf{x}},t)}\right]\,,
\end{equation}
with the functional derivatives $\delta/\delta X_i$, $\delta/\delta \Pi^i$ for the canonical variables and momenta, respectively.

The canonical Hamiltonian that results from the Legendre transformation of the Lagrange density given in \eqref{eq:theory-definition} was obtained in ref.~\cite{Reyes:2021cpx}. A wise approach is to explore the scalar $u$ as well as the scalar and purely spacelike sectors of $s^{\mu\nu}$ separately. The canonical momenta for each of these sectors then follow from eqs.~(\ref{eq:theory-einstein-hilbert}) -- (\ref{eq:theory-definition-sij}):
\begin{subequations}
\begin{align}
\label{eq:canonical-momentum-scalar-sector}
\pi^{ij}&:=\frac{\partial(\mathcal{L}^{(0)}+\mathcal{L}^{(u)})}{\partial \dot{q}_{ij}}=\frac{\sqrt{q}}{2\kappa}\left[(1-u)(K^{ij}-q^{ij}K)+\frac{1}{N}q^{ij}\mathcal{L}_mu\right]\,, \\[2ex]
\label{eq:canonical-momentum-timelike-sector}
p^{ij}&:=\frac{\partial(\mathcal{L}^{(0)}+\mathcal{L}_1^{(s)})}{\partial \dot{q}_{ij}}=\frac{\sqrt{q}}{2\kappa}\left[(1-s^{\mathbf{nn}})(K^{ij}-q^{ij}K)+\frac{1}{2N}q^{ij}\mathcal{L}_ms^{\mathbf{nn}}\right]\,, \\[2ex]
\label{eq:canonical-momentum-K-spacelike-sector}
P^{ij}&:=\frac{\partial(\mathcal{L}^{(0)}+\mathcal{L}^{(s)}_2)}{\partial \dot{q}_{ij}}=\frac{\sqrt{q}}{2\kappa}\left[K^{ij}-q^{ij}K-(s^{il}K_l^{\phantom{l}j}+s^{jl}K_l^{\phantom{l}i})-\frac{1}{2N}\mathcal{L}_ms^{ij}\right]\,.
\end{align}
\end{subequations}
For the $u$, $s^{\mathbf{nn}}$, and $s^{ij}$ sector each, the canonical Hamiltonians have the form
\begin{subequations}
\label{eq:canonical-hamiltonians}
\begin{align}
\label{eq:hamiltonian-u}
{H_u}&=\int_{\Sigma_t}\mathrm{d}^3x\,\left[-\frac{\sqrt{q}}{2\kappa}N\left((1-u)R+2D^iD_iu\right)+\frac{\mathcal{L}_mu}{1-u}\left(\pi-\frac{3}{4}\frac{\sqrt{q}}{\kappa N}\mathcal{L}_mu\right)\right. \notag \\
&\phantom{{}={}}\hspace{1.4cm}\left.{}+\frac{2\kappa N}{\sqrt{q}(1-u)}\left(\pi^{ij}\pi_{ij}-\frac{\pi^2}{2}\right)-2(D_i\pi^{ij})N_j\right]\,, \displaybreak[0]\\[2ex]
\label{eq:hamiltonian-snn}
H_1&=\int_{\Sigma_t}\mathrm{d}^3x\,\left[-\frac{\sqrt{q}}{2\kappa}N(R+D^iD_is^{\mathbf{nn}})+\frac{\mathcal{L}_ms^{\mathbf{nn}}}{2(1-s^{\mathbf{nn}})}\left(p-\frac{3}{8}\frac{\sqrt{q}}{\kappa N}\mathcal{L}_ms^{\mathbf{nn}}\right)\right. \notag \\
&\phantom{{}={}}\hspace{1.4cm}\left.{}+\frac{2\kappa N}{\sqrt{q}(1-s^{\mathbf{nn}})}\left(p^{ij}p_{ij}-\frac{p^2}{2}\right)-2(D_i p^{ij})N_j\right]\,, \displaybreak[0]\\[2ex]
\label{eq:hamiltonian-sij}
H_2&=\int_{\Sigma_t}\mathrm{d}^3x\,\bigg[-\frac{\sqrt{q}}{2\kappa}N\left(R+s^{ij}R_{ij}-D_jD_is^{ij}\right)+\left(P_{ij}-\frac{P}{2}q_{ij}\right)\mathcal{L}_ms^{ij}\notag \\
&\phantom{{}={}}\hspace{1.5cm}+\frac{2\kappa N}{\sqrt{q}}\left(P^{ij}P_{ij}-(1-s^i_{\phantom{i}i})\frac{P^2}{2}-2s^{ij}(P_{ij}P-P_i^{\phantom{i}k}P_{kj})\right) \notag \\
&\phantom{{}={}}\hspace{1.5cm}-2(D_i P^{ij})N_j\bigg]+\mathcal{O}[(s^{ij})^2]\,.
\end{align}
\end{subequations}
Note that $H_2$ is valid at first order in the coefficients $s^{ij}$. Since the latter parameterize a modification of GR, the coefficients are expected to be $\ll 1$ --- at least in regions with weak gravitational fields. Studying contributions to $H_2$ that are of higher order in $s^{ij}$ may be worthwhile, but it is very reasonable to understand the properties of this modification at leading order first.
%................................................................................
\section{Field equations in the hypersurface}
\label{eq:projected-field-equations}
%................................................................................
Coordinate invariance allows us to work in coordinates with a nontrivial lapse function, $N=N(t,\mathbf{x})$, and a zero shift vector, $N^i=0$, without a loss of generality.
In the following, we will focus on the $u$ sector first and set $s^{\mu\nu}=0$. Later on, we will be studying the complementary $s^{\mu\nu}$ sector with $u=0$. Now, the modified Einstein equations for a nonzero $u$ read \cite{Bailey:2006fd}
\begin{equation}
\label{eq:modified-einstein-u}
Q^{\mu\nu}:=(1-u){}^{(4)}G^{\mu\nu}+\nabla^{\mu}\nabla^{\nu}u-g^{\mu\nu}\square u=0\,.
\end{equation}
Their projection into the hypersurface $\Sigma_t$ is obtained from
\begin{equation}
(1-u)q^i_{\phantom{i}\mu}q^j_{\phantom{j}\nu}{}^{(4)}G^{\mu\nu}+q^i_{\phantom{i}\mu}q^j_{\phantom{j}\nu}\nabla^{\mu}\nabla^{\nu}u-q^{ij}\Box u=0\,.
\end{equation}
Carrying out the projection of the part proportional to the Einstein tensor is straightforward and the result is readily found to be
\begin{align}
\label{eq:einstein-tensor-projection}
q^i_{\phantom{i}\mu}q^j_{\phantom{j}\nu}{}^{(4)}G^{\mu\nu}&=R^{ij}-\frac{R}{2}q^{ij}+q^{ia}q^{jb}\frac{1}{N} \mathcal{L}_m K_{ab}
-\frac{1}{N}D^iD^jN+KK^{ij}-2K^{ik}K_k^{\phantom{k}j} \notag \\
&\phantom{{}={}}-\frac{1}{2} q^{ij} \left(K^2+K_{kl}K^{kl}+\frac{2}{N}\mathcal{L}_m K-\frac{2}{N}D_{l}D^{l}N\right)\,.
\end{align}
The covariant-derivative pieces are projected in the following way. Consider
\begin{equation}
q^i_{\phantom{i}\mu}q^j_{\phantom{j}\nu}\nabla^{\mu}\nabla^{\nu}u=D^iD^ju-\frac{1}{N}K^{ij}\mathcal{L}_mu\,,
\end{equation}
as well as
\begin{equation}
\square u=D^2u-\frac{1}{N}\left[K\mathcal{L}_m u+\mathcal{L}_m\left(\frac{1}{N}\mathcal{L}_mu\right)\right]+a^{\mu}\nabla_{\mu}u\,.
\end{equation}
We then intend to express the projected field equations in terms of the canonical momentum density $\pi^{ij}$ that is given by \eqref{eq:canonical-momentum-scalar-sector}. Inverting the latter for the extrinsic-curvature tensor $K^{ij}$ gives
\begin{equation}
\label{eq:extrinsic-curvature-tensor-u}
K^{ij}=\frac{1}{1-u}\left[\frac{2\kappa}{\sqrt{q}}\left(\pi^{ij}-\frac{\pi}{2}q^{ij}\right)+\frac{q^{ij}}{2N}\mathcal{L}_mu\right]\,.
\end{equation}
Taking into consideration the previous ingredients, the left-hand side of the modified Einstein equations completely projected into $\Sigma_t$ reads
\begin{subequations}
\label{eq:projected-einstein-equation-u}
\begin{align}
(\vec{\boldsymbol{q}}^{*}\mathbf{Q})^{ij}&=\frac{2\kappa}{N\sqrt{q}}\dot{\pi}^{ij}+(1-u)\left(R^{ij}-\frac{R}{2}q^{ij}\right) \notag \\
&\phantom{{}={}}+\frac{1}{N}\left(q^{ij}D_kD^k[(1-u)N]-D^iD^j[(1-u)N]\right) \notag \\
&\phantom{{}={}}+q^{ij}a^kD_ku-(a^iD^ju+a^jD^iu) \notag \\
&\phantom{{}={}}+\frac{4\kappa^2}{q(1-u)}\left[2\pi^{ik}\pi_k^{\phantom{k}j}-\pi\pi^{ij}-\frac{1}{2}\left(\pi_{kl}\pi^{kl}-\frac{\pi^2}{2}\right)q^{ij}\right] \notag \\
&\phantom{{}={}}+\frac{\mathcal{L}_mu}{N(1-u)}\left(\frac{2\kappa}{\sqrt{q}}\pi^{ij}-\frac{3}{4N}q^{ij}\mathcal{L}_mu\right)\,,
\end{align}
with the time derivative of the canonical momentum stated in \eqref{eq:canonical-momentum-scalar-sector}:
\begin{align}
\dot{\pi}^{ij}&=\frac{\sqrt{q}}{2\kappa}\bigg[N(1-u)K(3K^{ij}-q^{ij}K)+(1-u)(\mathcal{L}_mK^{ij}-q^{ij}\mathcal{L}_mK) \notag \\
&\phantom{{}={}}\hspace{0.8cm}+(2q^{ij}K-3K^{ij})\mathcal{L}_mu+q^{ij}\mathcal{L}_m\left(\frac{1}{N}\mathcal{L}_mu\right)\bigg]\,.
\end{align}
\end{subequations}
Here, we took over the convenient notation from refs.~\cite{Gourgoulhon:2007ue,Gourgoulhon:2012} describing the projection of a generic spacetime tensor into $\Sigma_t$:
\begin{equation}
(\vec{\boldsymbol{q}}^{*}\mathbf{T})^{\alpha_1\dots\alpha_s}_{\phantom{\alpha_1\dots\alpha_s}\beta_1\dots\beta_t}:=q^{\alpha_1}_{\phantom{\alpha_1}\gamma_1}\dots q^{\alpha_s}_{\phantom{\alpha_s}\gamma_s}q^{\delta_1}_{\phantom{\delta_1}\beta_1}\dots q^{\delta_t}_{\phantom{\delta_t}\beta_t}T^{\gamma_1\dots\gamma_s}_{\phantom{\gamma_1\dots\gamma_s}\delta_1\dots\delta_t}\,.
\end{equation}
The next step is to consider the modified Einstein equations for the $s^{\mu\nu}$ coefficients~\cite{Bailey:2006fd}:
\begin{equation}
\label{eq:modified-einstein-s}
J^{\mu\nu}:={}^{(4)}G^{\mu\nu}-\frac{1}{2}s^{\alpha\beta}R_{\alpha\beta}g^{\mu\nu}-\frac{1}{2}\nabla_{\alpha}\nabla^{\mu}s^{\alpha\nu}-\frac{1}{2}\nabla_{\alpha}\nabla^{\nu}s^{\alpha\mu}+\frac{1}{2}\square s^{\mu\nu}+\frac{1}{2}g^{\mu\nu}\nabla_{\alpha}\nabla_{\beta}s^{\alpha\beta}=0\,.
\end{equation}
The projection of the Einstein tensor is quickly obtained from \eqref{eq:einstein-tensor-projection}. As we remarked before, we distinguish between the different sectors of $s^{\mu\nu}$ mentioned below \eqref{eq:decomposition-s}. Recall that the coefficients $s^{i\mathbf{n}}$ are pure gauge degrees of freedom, which is why they were discarded right from the start. To compute the projection of \eqref{eq:modified-einstein-s} for the purely timelike sector composed of the single scalar coefficient $s^{\mathbf{nn}}$, the following results are indispensable:
\begin{subequations}
\begin{align}
q^{ij}s^{\mathbf{nn}}R_{\mathbf{nn}}&=q^{ij}s^{\mathbf{nn}}\left(-\frac{1}{N}\mathcal{L}_mK+D_la^l+a_la^l-K^{kl}K_{kl}\right)\,, \displaybreak[0]\\[2ex]
q^i_{\phantom{i}\mu}q^j_{\phantom{j}\nu}\nabla_{\alpha}\nabla^{\mu}(s^{\mathbf{nn}}n^{\alpha}n^{\nu})&=(D^is^{\mathbf{nn}})a^j-s^{\mathbf{nn}}a^ia^j+s^{\mathbf{nn}}(KK^{ij}-K^{ki}K^j_{\phantom{j}k}) \notag \\
&\phantom{{}={}}+\frac{1}{N}(\mathcal{L}_ms^{\mathbf{nn}})K^{ij}+\frac{s^{\mathbf{nn}}}{N}q^{ik}q^{jl}\mathcal{L}_mK_{kl}\,, \displaybreak[0]\\[2ex]
q^i_{\phantom{i}\alpha}q^j_{\phantom{j}\beta}\square(s^{\mathbf{nn}}n^{\mu}n^{\nu})&=2s^{\mathbf{nn}}(K^{ik}K^j_{\phantom{j}k}-a^ia^j)\,, \displaybreak[0]\\[2ex]
q^i_{\phantom{i}\mu}q^j_{\phantom{j}\nu}(g^{\mu\nu}\nabla_{\alpha}\nabla_{\beta}s^{\alpha\beta})&=q^{ij}\left[\frac{1}{N}\mathcal{L}_m\left(\frac{1}{N}\mathcal{L}_ms^{\mathbf{nn}}\right)+a^kD_ks^{\mathbf{nn}}+\frac{2}{N}(\mathcal{L}_ms^{\mathbf{nn}})K\right. \notag \\
&\phantom{{}={}}\hspace{0.5cm}\left.{}+s^{\mathbf{nn}}\left(\frac{1}{N}\mathcal{L}_mK+K^2\right)+s^{\mathbf{nn}}(D_la^l+a_la^l)\right]\,.
\end{align}
\end{subequations}
Here, the extrinsic curvature expressed in terms of the canonical momentum reads
\begin{equation}
\label{eq:extrinsic-curvature-tensor-snn}
K^{ij}=\frac{1}{1-s^{\mathbf{nn}}}\left[\frac{2\kappa}{\sqrt{q}}\left(p^{ij}-\frac{p}{2}q^{ij}\right)+\frac{q^{ij}}{4N}\mathcal{L}_ms^{\mathbf{nn}}\right]\,.
\end{equation}
The projected field equations are simpler in comparison to those of the $u$ sector, as there is no term of the form $s^{\mathbf{nn}}R$ (cf.~eqs.~(\ref{eq:theory-definition-u}), (\ref{eq:theory-definition-u})). Now we are in a position to evaluate the left-hand side of \eqref{eq:modified-einstein-s} for $s^{i\mathbf{n}}=s^{ij}=0$ completely projected into $\Sigma_t$. It is useful to define $(J_1)^{\mu\nu}:=J^{\mu\nu}|_{s^{i\mathbf{n}}=s^{ij}=0}$ with $J^{\mu\nu}$ given in \eqref{eq:modified-einstein-s}. Then,
\begin{subequations}
\label{eq:projected-einstein-equation-snn}
\begin{align}
(\vec{\boldsymbol{q}}^{*}\mathbf{J}_1)^{ij}&=\frac{2\kappa}{N\sqrt{q}}\dot{p}^{ij}+R^{ij}-\frac{R}{2}q^{ij}+\frac{1}{N}(q^{ij}D_kD^kN-D^iD^jN) \notag \\
&\phantom{{}={}}+\frac{1}{2}\left[q^{ij}a^kD_ks^{\mathbf{nn}}-(a^iD^js^{\mathbf{nn}}+a^jD^is^{\mathbf{nn}})\right] \notag \\
&\phantom{{}={}}+\frac{4\kappa^2}{q(1-s^{\mathbf{nn}})}\left[2p^{ik}p_k^{\phantom{k}j}-pp^{ij}-\frac{1}{2}\left(p_{kl}p^{kl}-\frac{p^2}{2}\right)q^{ij}\right] \notag \\
&\phantom{{}={}}+\frac{\mathcal{L}_ms^{\mathbf{nn}}}{N(1-s^{\mathbf{nn}})}\left(\frac{\kappa}{\sqrt{q}}\pi^{ij}-\frac{3}{16N}q^{ij}\mathcal{L}_ms^{\mathbf{nn}}\right)\,,
\end{align}
with the following time derivative of the canonical momentum of \eqref{eq:canonical-momentum-timelike-sector}:
\begin{align}
\dot{p}^{ij}&=\frac{\sqrt{q}}{2\kappa}\bigg[N(1-s^{\mathbf{nn}})K(3K^{ij}-q^{ij}K)+(1-s^{\mathbf{nn}})(\mathcal{L}_mK^{ij}-q^{ij}\mathcal{L}_mK) \notag \\
&\phantom{{}={}}\hspace{0.7cm}+\left(\frac{3}{2}q^{ij}K-2K^{ij}\right)\mathcal{L}_ms^{\mathbf{nn}}+\frac{q^{ij}}{2}\mathcal{L}_m\left(\frac{1}{N}\mathcal{L}_ms^{\mathbf{nn}}\right)\bigg]\,.
\end{align}
\end{subequations}
Finally, we turn to the $s^{ij}$ sector. We need
\begin{subequations}
\label{eq:intermediate-results-sij}
\begin{align}
q^i_{\phantom{i}\alpha}q^j_{\phantom{j}\beta}g^{\alpha\beta}s^{\mu\nu}R_{\mu\nu}&=q^{ij}s^{kl}\bigg[\frac{1}{N}\mathcal{L}_mK_{kl}-(D_ka_l+a_ka_l)+R_{kl} \notag \\
&\phantom{{}={}}\hspace{1.1cm}+KK_{kl}-2K_{kn}K^n_{\phantom{n}l}\bigg]\,, \displaybreak[0]\\[2ex]
q^i_{\phantom{i}\alpha}q^j_{\phantom{j}\beta}\nabla_{\nu}\nabla^{\alpha}(q^{\nu}_{\phantom{\nu}\gamma}q^{\beta}_{\phantom{\beta}\delta}s^{\gamma\delta})&=KK^i_{\phantom{i}k}s^{kj}+2K^i_{\phantom{i}k}K^k_{\phantom{k}l}s^{lj}+D_kD^is^{kj}-a^ia_ks^{kj}+a_kD^is^{kj} \notag \\
&\phantom{{}={}}+\frac{1}{N}(\mathcal{L}_mK^{il})q_{lk}s^{kj}+K^i_{\phantom{i}k}s^{kl}K_l^{\phantom{l}j}\,, \displaybreak[0]\\[2ex]
q^i_{\phantom{i}\alpha}q^j_{\phantom{j}\beta}\square(q^{\alpha}_{\phantom{\alpha}\gamma}q^{\beta}_{\phantom{\beta}\delta}s^{\gamma\delta})&=D_kD^ks^{ij}-K\left(\frac{1}{N}\mathcal{L}_ms^{ij}+s^{ik}K_k^{\phantom{k}j}+s^{jk}K_k^{\phantom{k}i}\right) \notag \\
&\phantom{{}={}}+K^i_{\phantom{i}k}K^k_{\phantom{k}l}s^{lj}+K^j_{\phantom{j}k}K^k_{\phantom{k}l}s^{il} \notag \displaybreak[0] \\
&\phantom{{}={}}-\bigg[\frac{1}{N}\mathcal{L}_m\left(\frac{1}{N}\mathcal{L}_ms^{ij}\right)+\frac{1}{N}\left(K_k^{\phantom{k}j}\mathcal{L}_ms^{ik}+K^i_{\phantom{i}k}\mathcal{L}_ms^{kj}\right) \notag \\
&\phantom{{}={}}\hspace{0.5cm}+\left(\frac{1}{N}\mathcal{L}_ms^{ik}+s^{il}K_l^{\phantom{l}k}+s^{kl}K_l^{\phantom{l}i}\right)K_k^{\phantom{k}j} \notag \displaybreak[0]\\
&\phantom{{}={}}\hspace{0.5cm}+s^{ik}q_{kl}\left(\frac{1}{N}\mathcal{L}_mK^{lj}+2K^{lm}K_m^{\phantom{m}j}\right) \notag \displaybreak[0]\\
&\phantom{{}={}}\hspace{0.5cm}+\left(\frac{1}{N}\mathcal{L}_ms^{jk}+s^{jl}K_l^{\phantom{l}k}+s^{kl}K_l^{\phantom{l}j}\right)K_k^{\phantom{k}i} \notag \displaybreak[0]\\
&\phantom{{}={}}\hspace{0.5cm}+s^{jk}q_{kl}\left(\frac{1}{N}\mathcal{L}_mK^{li}+2K^{lm}K_m^{\phantom{m}i}\right) \notag \displaybreak[0]\\
&\phantom{{}={}}\hspace{0.5cm}-a_kD^ks^{ij}+a^is^{jk}a_k+a^js^{ik}a_k\bigg]\,, \displaybreak[0]\\[2ex]
q^i_{\phantom{i}\alpha}q^j_{\phantom{j}\beta}(g^{\alpha\beta}\nabla_{\mu}\nabla_{\nu}s^{\mu\nu})&=q^{ij}\bigg[\frac{1}{N}\mathcal{L}_m(K_{kl}s^{kl})+s^{kl}D_ka_l+2a_kD_ls^{kl} \notag \\
&\phantom{{}={}}\hspace{0.6cm}+a_ia_js^{ij}+D_kD_ls^{kl}+KK_{kl}s^{kl}\bigg]\,.
\end{align}
\end{subequations}
Furthermore, the extrinsic-curvature tensor is given by
\begin{align}
\label{eq:extrinsic-curvature-tensor-sij}
K^{ij}&=\frac{2\kappa}{\sqrt{q}}\left(P^{ij}-\frac{1}{2}\left[(1-s^k_{\phantom{k}k})P+2s^{kl}P_{kl}\right]q^{ij}+s^{ik}P_k^{\phantom{k}j}+s^{jk}P_k^{\phantom{k}i}-s^{ij}P\right) \notag \\
&\phantom{{}={}}+\frac{1}{2N}\left(\mathcal{L}_ms^{ij}-\frac{q^{ij}}{2}q_{kl}\mathcal{L}_ms^{kl}\right)+\mathcal{O}[(s^{ij})^2]\,.
\end{align}
Let $(J_2)^{\mu\nu}:=J^{\mu\nu}|_{s^{\mathbf{nn}}=s^{i\mathbf{n}}=0}$ be the left-hand side of the modified Einstein equations of \eqref{eq:modified-einstein-s} for the purely spacelike sector composed of the coefficients $s^{ij}$. Considering their projection into $\Sigma_t$ and incorporating \eqref{eq:intermediate-results-sij} as well as expressing the extrinsic curvature in terms of the canonical momentum according to \eqref{eq:canonical-momentum-K-spacelike-sector} results in
\begin{subequations}
\label{eq:projected-einstein-equation-sij}
\begin{align}
(\vec{\boldsymbol{q}}^{*}\mathbf{J}_2)^{ij}&=\frac{2\kappa}{N\sqrt{q}}\dot{P}^{ij}+R^{ij}-\frac{R}{2}q^{ij}+\frac{1}{N}(q^{ij}D_kD^kN-D^iD^jN) \notag \displaybreak[0]\\
&\phantom{{}={}}+\frac{1}{2}\left[-q^{ij}(s^{kl}R_{kl}-D_kD_ls^{kl})-D_kD^is^{kj}-D_kD^js^{ki}+D_kD^ks^{ij}\right] \notag \displaybreak[0]\\
&\phantom{{}={}}+\frac{1}{2}\left(2q^{ij}\left[a_kD_ls^{kl}+s^{kl}(D_ka_l+a_ka_l)\right]-a_k(D^is^{kj}+D^js^{ki})+a_kD^ks^{ij}\right) \notag \displaybreak[0]\\
&\phantom{{}={}}+\frac{\kappa^2}{q}\Big(\left[(1-s^k_{\phantom{k}k})P^2-2P^{kl}P_{kl}+4s^{kl}(PP_{kl}-P_k^{\phantom{k}m}P_{lm})\right]q^{ij} \notag \displaybreak[0]\\
&\phantom{{}={}}\hspace{1.2cm}+8\bigg[P^{ik}P_k^{\phantom{k}j}+s^{ik}P_{kl}P^{lj}+s^{jk}P_{kl}P^{li}-s^{ik}PP_k^{\phantom{k}j}-s^{jk}PP_k^{\phantom{k}i} \notag \displaybreak[0]\\
&\phantom{{}={}}\hspace{1.2cm}+s^{kl}(P^i_{\phantom{i}k}P_l^{\phantom{l}j}-P_{kl}P^{ij})\Big]+2P^2s^{ij}-4(1-s^k_{\phantom{k}k})PP^{ij}\Big) \notag \displaybreak[0]\\
&\phantom{{}={}}+\frac{\kappa}{N\sqrt{q}}\left[2\left(P^i_{\phantom{i}k}\mathcal{L}_ms^{jk}+P^j_{\phantom{j}k}\mathcal{L}_ms^{ik}\right)-\left(P\mathcal{L}_ms^{ij}+P^{ij}q_{kl}\mathcal{L}_ms^{kl}\right)\right]\,,
\end{align}
where the time derivative of the canonical momentum of \eqref{eq:canonical-momentum-K-spacelike-sector} reads
\begin{align}
\dot{P}^{ij}&=\frac{\sqrt{q}}{2\kappa}\left[NK(3K^{ij}-q^{ij}K)-N(s^{il}K_l^{\phantom{l}j}+s^{jl}K_l^{\phantom{l}i})K+\mathcal{L}_mK^{ij}-(s^{il}\mathcal{L}_mK_l^{\phantom{l}j}+s^{jl}\mathcal{L}_mK_l^{\phantom{l}i})\right. \notag \\
&\phantom{{}={}}\hspace{0.6cm}\left.{}-q^{ij}\mathcal{L}_mK-\frac{K}{2}\mathcal{L}_ms^{ij}-(\mathcal{L}_ms^{il})K_l^{\phantom{l}j}-(\mathcal{L}_ms^{jl})K_l^{\phantom{l}i}-\frac{1}{2}\mathcal{L}_m\left(\frac{1}{N}\mathcal{L}_ms^{ij}\right)\right]\,.
\end{align}
\end{subequations}
With the latter result at hand, the modified Einstein equations for the scalar sectors $u$ and $s^{\mathbf{nn}}$ as well as the purely spacelike sector $s^{ij}$ have been successfully projected into spacelike hypersurfaces $\Sigma_t$.

%................................................................................
\section{Hamilton-Jacobi equations of motion}
\label{sec:hamilton-jacobi}
%................................................................................
With the projected altered Einstein equations of the modified-gravity theory based on \eqref{eq:theory-definition} at our disposal, in this section we will be working within the Hamiltonian formalism. The latter is based on the following set of field equations that are of first order in time:
\begin{subequations}
\begin{align}
\label{eq:hamilton-equation-1}
\dot{q}_{ij}&=\{q_{ij},H\}\,, \\[2ex]
\label{eq:hamilton-equation-2}
\dot{\pi}^{ij}&=\{\pi^{ij},H\}\,.
\end{align}
\end{subequations}
The first set of equations corresponds to a geometrical identity \cite{Gambini:1996}, whereas the second set describes the dynamics of the theory under consideration \cite{Bertschinger:2002}. In the latter, we employ the canonical Hamiltonians of the $u$, $s^{\mathbf{nn}}$, and $s^{ij}$ sectors given in \eqref{eq:canonical-hamiltonians}. Our goal is to compare the projected Einstein equations stated in eqs.~(\ref{eq:projected-einstein-equation-u}), (\ref{eq:projected-einstein-equation-snn}), and (\ref{eq:projected-einstein-equation-sij}) to \eqref{eq:hamilton-equation-2} evaluated for each of the canonical Hamiltonians of \eqref{eq:canonical-hamiltonians}.

The canonical Hamiltonian associated with the $u$ sector is given by \eqref{eq:hamiltonian-u}. First of all, we will investigate the first set of Hamilton's equations in \eqref{eq:hamilton-equation-1}, i.e.,
\begin{equation}
\dot{q}_{ij}=\{q_{ij},{H_u}\}\,.
\end{equation}
Evaluating the Poisson bracket leads to
\begin{equation}
\left\{q_{ij},\int_{\Sigma_t}\mathrm{d}^3x\,\frac{1}{1-u}\left[\pi\mathcal{L}_mu+\frac{2\kappa N}{\sqrt{q}}\left(\pi^{kl}\pi_{kl}-\frac{\pi^2}{2}\right)\right]\right\}=2NK_{ij}\,,
\end{equation}
where we used the extrinsic-curvature tensor of \eqref{eq:extrinsic-curvature-tensor-u} given as a function of the canonical momentum. After confirming the outcome expected for the Hamiltonian describing the gravitational system, we devote ourselves to the second set of Hamilton's equations in \eqref{eq:hamilton-equation-2}:
\begin{subequations}
\begin{align}
-\left\{\pi^{ij},\int_{\Sigma_t}\mathrm{d}^3x\,\frac{\sqrt{q}}{2\kappa}N(1-u)R\right\}&=-\frac{\sqrt{q}}{2\kappa}\Big(q^{ij}D_lD^l[(1-u)N]-D^iD^j[(1-u)N] \notag \\
&\phantom{{}={}}\hspace{1.1cm}+N(1-u)G^{ij}\Big)\,, \displaybreak[0]\\[2ex]
\left\{\pi^{ij},\int_{\Sigma_t}\mathrm{d}^3x\,\frac{2\kappa N}{\sqrt{q}(1-u)}\left(\pi^{ij}\pi_{ij}-\frac{\pi^2}{2}\right)\right\}&=-\frac{2\kappa N}{\sqrt{q}(1-u)}\bigg[2\pi^{ia}\pi^j_{\phantom{j}a}-\pi\pi^{ij} \notag \\
&\phantom{{}={}}\hspace{2.2cm}-\frac{1}{2}\left(\pi_{ab}\pi^{ab}-\frac{\pi^2}{2}\right)q^{ij}\bigg]\,, \displaybreak[0]\\[2ex]
-\left\{\pi^{ij},\int_{\Sigma_t}\mathrm{d}^3x\,\frac{N\sqrt{q}}{\kappa}D^iD_iu\right\}&=-\frac{N\sqrt{q}}{2\kappa}q^{ij}a_lD^lu\,, \displaybreak[0]\\[2ex]
\left\{\pi^{ij},\int_{\Sigma_t}\mathrm{d}^3x\,\frac{\mathcal{L}_mu}{1-u}\left(\pi-\frac{3}{4}\frac{\sqrt{q}}{\kappa N}\mathcal{L}_mu\right)\right\}&=-\frac{\mathcal{L}_mu}{1-u}\left(\pi^{ij}-\frac{3}{8}\frac{\sqrt{q}}{\kappa N}q^{ij}\mathcal{L}_mu\right)\,,
\end{align}
\end{subequations}
with the Einstein tensor $G^{ij}$ in $\Sigma_t$. Organizing the individual parts, a multiplication by $2\kappa/(N\sqrt{q})$ reproduces \eqref{eq:projected-einstein-equation-u} apart from the last two terms in the third line, which depend on the acceleration. The boundary term of \eqref{eq:additional-boundary-term} solves this issue. The part proportional to $u$ is interpreted as an additional contribution to ${H_u}$ of \eqref{eq:hamiltonian-u} that contains a total derivative:
\begin{equation}
\Delta_u:=-\int_{\Sigma_t}\mathrm{d}^3x\,\frac{\sqrt{q}}{\kappa}D_l(ND^lu)=-\int_{\Sigma_t}\mathrm{d}^3x\,\frac{N\sqrt{q}}{\kappa}(D_lD^lu+a_lD^lu)\,.
\end{equation}
Its Poisson bracket with the canonical momentum density reads
\begin{equation}
\{\pi^{ij},\Delta_u\}=\frac{N\sqrt{q}}{2\kappa}(a^iD^ju+a^jD^iu)\,,
\end{equation}
and provides the missing pieces to \eqref{eq:projected-einstein-equation-u}.

Next, our focus is on the $s^{\mathbf{nn}}$ and $s^{ij}$ sectors. A computation reveals that the first set of Hamilton's equations in \eqref{eq:hamilton-equation-1} is satisfied for these sectors, too. Thus, for the Hamiltonians $H_1$ and $H_2$ of \eqref{eq:hamiltonian-snn} and \eqref{eq:hamiltonian-sij}, respectively, it holds that
\begin{subequations}
\begin{align}
\dot{q}_{ij}&=\{q_{ij},H_1\}=\left\{q_{ij},\int_{\Sigma_t}\mathrm{d}^3x\,\frac{1}{1-s^{\mathbf{nn}}}\left[\frac{p}{2}\mathcal{L}_ms^{\mathbf{nn}}+\frac{2\kappa N}{\sqrt{q}}\left(p^{kl}p_{kl}-\frac{p^2}{2}\right)\right]\right\}=2NK_{ij}\,, \displaybreak[0]\\[2ex]
\dot{q}_{ij}&=\{q_{ij},H_2\}=\left\{q_{ij},\int_{\Sigma_t}\mathrm{d}^3x\,\left[\frac{2\kappa N}{\sqrt{q}}\left(P^{kl}P_{kl}-(1-s^k_{\phantom{k}k})\frac{P^2}{2}-2s^{kl}(P_{kl}P-P_k^{\phantom{k}n}P_{nl})\right)\right.\right. \notag \\
&\phantom{{}={}}\left.\left.\hspace{4.2cm}{}+\left(P_{kl}-\frac{P}{2}q_{kl}\right)\mathcal{L}_ms^{kl}\right]\right\}=2NK_{ij}\,,
\end{align}
\end{subequations}
where the right-hand sides are given by the extrinsic-curvature tensors of \eqref{eq:extrinsic-curvature-tensor-snn} and \eqref{eq:extrinsic-curvature-tensor-sij}, respectively. With that settled, we take a look at the second set of Hamilton's equations of \eqref{eq:hamilton-equation-2} for the $s^{\mathbf{nn}}$ sector. The individual contributions to the Poisson bracket are
\begin{subequations}
\begin{align}
-\left\{p^{ij},\int_{\Sigma_t}\mathrm{d}^3x\,\frac{\sqrt{q}}{2\kappa}NR\right\}&=-\frac{\sqrt{q}}{2\kappa}\left(q^{ij}D_lD^lN-D^iD^jN+NG^{ij}\right)\,, \displaybreak[0]\\[2ex]
\left\{p^{ij},\int_{\Sigma_t} \mathrm{d}^3x\,\frac{2\kappa N}{\sqrt{q}(1-s^{\mathbf{nn}})}\left(p^{ij}p_{ij}-\frac{p^2}{2}\right)\right\}&=-\frac{2\kappa N}{\sqrt{q}(1-s^{\mathbf{nn}})}\bigg[2p^{ia}p^j_{\phantom{j}a}-pp^{ij} \notag \\
&\phantom{{}={}}\hspace{2.6cm}-\frac{1}{2}\left(p_{ab}p^{ab}-\frac{p^2}{2}\right)q^{ij}\bigg]\,, \displaybreak[0]\\[2ex]
-\left\{p^{ij},\int_{\Sigma_t}\mathrm{d}^3x\,\frac{N\sqrt{q}}{2\kappa}D^iD_is^{\mathbf{nn}}\right\}&=-\frac{N\sqrt{q}}{4\kappa}q^{ij}a_lD^ls^{\mathbf{nn}}\,, \displaybreak[0]\\[2ex]
\left\{p^{ij},\int_{\Sigma_t}\mathrm{d}^3x\,\frac{\mathcal{L}_ms^{\mathbf{nn}}}{2(1-s^{\mathbf{nn}})}\left(p-\frac{3}{8}\frac{\sqrt{q}}{\kappa N}\mathcal{L}_ms^{\mathbf{nn}}\right)\right\}&=-\frac{\mathcal{L}_ms^{\mathbf{nn}}}{2(1-s^{\mathbf{nn}})}\left(p^{ij}-\frac{3}{16}\frac{\sqrt{q}}{\kappa N}q^{ij}\mathcal{L}_ms^{\mathbf{nn}}\right)\,.
\end{align}
\end{subequations}
By using these findings and multiplying by $2\kappa/(N\sqrt{q})$, we arrive at \eqref{eq:projected-einstein-equation-snn} with the exception of the last two terms in the second line depending on the acceleration. Recall that an analogous issue occurred previously in the $u$ sector. The boundary term of \eqref{eq:additional-boundary-term} again solves with problem. The part proportional to $s^{\mathbf{nn}}$ contributes to $H_1$ of \eqref{eq:hamiltonian-snn} with a total derivative:
\begin{equation}
\Delta_1:=-\int_{\Sigma_t}\mathrm{d}^3x\,\frac{\sqrt{q}}{2\kappa}D_l(ND^ls^{\mathbf{nn}})=-\int_{\Sigma_t}\mathrm{d}^3x\,\frac{N\sqrt{q}}{2\kappa}(D_lD^ls^{\mathbf{nn}}+a_lD^ls^{\mathbf{nn}})\,.
\end{equation}
Computing the Poisson bracket of the latter contribution with the canonical momentum density results in
\begin{equation}
\{\pi^{ij},\Delta_1\}=\frac{N\sqrt{q}}{4\kappa}(a^iD^js^{\mathbf{nn}}+a^jD^is^{\mathbf{nn}})\,,
\end{equation}
which is related to the missing parts in \eqref{eq:projected-einstein-equation-snn}.

Finally, the most intricate sector of the $s^{ij}$ coefficients is on the menu. Here, we benefit from the results
\begin{subequations}
\label{eq:poisson-brackets-sij}
\begin{align}
-\left\{P^{ij},\int_{\Sigma_t}\mathrm{d}^3x\,\frac{\sqrt{q}}{2\kappa}NR\right\}&=-\frac{\sqrt{q}}{2\kappa}\left(NG^{ij}+q^{ij}D_lD^lN-D^iD^jN\right)\,, \displaybreak[0]\\[2ex]
-\left\{P^{ij},\int_{\Sigma_t}\mathrm{d}^3x\,\frac{N\sqrt{q}}{2\kappa}(s^{kl}R_{kl}-D_lD_ks^{kl})\right\}&=\frac{N\sqrt{q}}{4\kappa}\Big(-2q^{ij}\left[a_kD_ls^{kl}+s^{kl}(D_ka_l+a_ka_l)\right] \notag \\
&\phantom{{}={}}\hspace{1.2cm}+a_k(D^is^{kj}+D^js^{ki})-a_kD^ks^{ij} \notag \\
&\phantom{{}={}}\hspace{1.2cm}+q^{ij}(s^{kl}R_{kl}-D_kD_ls^{kl})+D_kD^is^{kj} \notag \\
&\phantom{{}={}}\hspace{1.2cm}+D_kD^js^{ki}-D_kD^ks^{ij}\Big)\,, \displaybreak[0]\\[2ex]
\left\{P^{ij},\int_{\Sigma_t}\mathrm{d}^3x\,\left(P_{kl}-\frac{P}{2}q_{kl}\right)\mathcal{L}_ms^{kl}\right\}&=\frac{1}{2}(P^{ij}q_{kl}\mathcal{L}_ms^{kl}+P\mathcal{L}_ms^{ij}) \notag \\
&\phantom{{}={}}-(P^i_{\phantom{i}k}\mathcal{L}_ms^{kj}+P^j_{\phantom{j}k}\mathcal{L}_ms^{ki})\,,
\end{align}
as well as
\begin{align}
&\left\{P^{ij},\int_{\Sigma_t}\mathrm{d}^3x\,\frac{2\kappa N}{\sqrt{q}}\left[P^{ij}P_{ij}-\frac{1}{2}(1-s^k_{\phantom{k}k})P^2-2s^{kl}(PP_{kl}-P_k^{\phantom{k}n}P_{ln})\right]\right\} \notag \displaybreak[0]\\
&=-\frac{\kappa N}{2\sqrt{q}}\Big(\left[(1-s^k_{\phantom{k}k})P^2-2P^{kl}P_{kl}+4s^{kl}(PP_{kl}-P_k^{\phantom{k}m}P_{lm})\right]q^{ij} \notag \displaybreak[0]\\
&\phantom{{}={}}\hspace{1.3cm}+8\Big[P^{ik}P_k^{\phantom{k}j}+s^{ik}P_{kl}P^{lj}+s^{jk}P_{kl}P^{li}-s^{ik}PP_k^{\phantom{k}j}-s^{jk}PP_k^{\phantom{k}i} \notag \displaybreak[0]\\
&\phantom{{}={}}\hspace{1.3cm}+s^{kl}(P^i_{\phantom{i}k}P_l^{\phantom{l}j}-P_{kl}P^{ij})\Big]\Big)+2P^2s^{ij}-4(1-s^k_{\phantom{k}k})PP^{ij}\,.
\end{align}
\end{subequations}
We employ the latter findings and multiply by $2\kappa/(N\sqrt{q})$, which yields \eqref{eq:projected-einstein-equation-sij}. Hence, our conclusion is that the Lagrangian and Hamiltonian formulation of \eqref{eq:theory-definition} lead to the same dynamics. The lengthy computations necessary to demonstrate this outcome corroborate the nontrivial nature of this property.

\section{Outlook: Other modified-gravity theories and the SME}
\label{sec:outlook}

After demonstrating the dynamical consistency of \eqref{eq:minimal-gravity-sme-reformulated}, two additional correspondences between modified-gravity theories and particular sectors of the gravitational SME shall be presented. By doing so, we intend to foster further interest in SME-related physics.

\subsection{Linearized dRGT massive gravity}
\label{eq:linearized-dRGT-theory}

The recent paper~\cite{Kostelecky:2021xhb} is dedicated to gaining a better theoretical understanding of dRGT theory with an emphasis on the stability of static solutions as well as certain properties of the linearized theory such as gravitational-wave propagation and the gravitational energy between two pointlike masses. To be able to study the propagation of gravitational-wave modes, the linearized field equations are necessary. In momentum space, they are cast into the form
\begin{equation}
\tilde{O}_{\mu\nu}^{\phantom{\mu\nu}\alpha\beta}\tilde{h}_{\alpha\beta}=0\,,
\end{equation}
where $\tilde{h}_{\alpha\beta}$ corresponds to the Fourier transform of the dynamical-metric perturbation $h_{\mu\nu}:=g_{\mu\nu}-\eta_{\mu\nu}$ and $\tilde{O}_{\mu\nu}^{\phantom{\mu\nu}\alpha\beta}$ is a tensor-valued operator in momentum space given by their eq.~(82). Furthermore, the authors describe the reference metric via a small deviation from the frequently taken Minkowski metric $\eta_{\mu\nu}$, i.e., they define $\delta f^{\mu\nu}:=f^{\mu\nu}-\eta^{\mu\nu}\ll 1$. All results are stated at first order in the perturbation $\delta f^{\mu\nu}$.

The modifications of GR are proportional to five dimensionless parameters $c_i$ with $i\in \{1\dots 5\}$. Now, all possible background fields with an arbitrary mass dimension preserving or violating diffeomorphism invariance in linearized gravity are classified in ref.~\cite{Kostelecky:2017zob}. Appropriate background fields violating diffeomorphism invariance can be associated with each of these contributions via their symmetry properties and momentum dependences. The only suitable coefficients are those of mass dimension 2 that are totally symmetric and not contracted with additional momenta: $k^{(2,1)\mu\nu\varrho\sigma}$. For these SME coefficients we refer to line 11 in Table~1 of ref.~\cite{Kostelecky:2017zob}. Then,
\begin{align}
k^{(2,1)\mu\nu\varrho\sigma}&=-\frac{m^2}{2}\bigg(c_1(\eta^{\varrho(\mu}\eta^{\nu)\sigma}-\eta^{\mu\nu}\eta^{\varrho\sigma})+c_2\eta^{(\varrho\mu}\eta^{\sigma\nu)}+c_3(\eta^{-1}\delta f)\eta^{\varrho(\mu}\eta^{\nu)\sigma} \notag \\
&\phantom{{}={}}\hspace{1.1cm}+c_4\left[\delta f^{\mu\nu}-\eta^{\mu\nu}(\eta^{-1}\delta f)\right]\eta^{\varrho\sigma}+c_5\eta^{\mu\nu}\delta f^{\varrho\sigma}\bigg)\,,
\end{align}
where $m$ is the graviton mass and indices enclosed by a pair of parentheses indicate that a symmetrization is performed in these indices. The latter coefficients characterize the linear regime of dRGT massive gravity completely.

\subsection{Ho\v{r}ava-Lifshitz gravity}
\label{sec:horava-lifshitz-gravity}

Ho\v{r}ava-Lifshitz gravity \cite{Horava:2008ih,Horava:2009uw} is a possible UV completion of GR. Its formulation relies on a foliation of spacetime into purely spacelike hypersurfaces via the ADM decomposition, while the leading contributions read~\cite{Blas:2009qj}
\begin{equation}
\label{eq:action-horava-lifshitz-gravity}
S_{\mathrm{HL}}=\frac{1}{16\pi G_H}\int_{\mathcal{M}}\mathrm{d}t\mathrm{d}^3x\,N\sqrt{q}(\xi R+K_{ij}K^{ij}-\lambda K^2+\alpha a_ia^i)\,,
\end{equation}
with the extrinsic curvature $K_{ij}$ and the acceleration $a_i$ defined directly under \eqref{eq:decompositions-ricci}. Furthermore, the action involves dimensionless dynamical parameters $\xi,\lambda,\alpha$ as well as the dimensionful gravitational constant $G_H$ of this theory. A significant property of Ho\v{r}ava-Lifshitz gravity are different weight factors associated with the contributions $K_{ij}K^{ij}$ and $K^2$. As it turns out, it is this property that makes a match to the $u$ and $s^{\mu\nu}$ sectors of \eqref{eq:theory-definition} presumably impossible. The authors of ref.~\cite{ONeal-Ault:2020ebv} also observed this issue, which is why they added further would-be dynamical terms to their action to enable a match. An alternative is to match \eqref{eq:action-horava-lifshitz-gravity} with a setting known as Einstein-aether theory \cite{Jacobson:2004ts,Elliott:2005va,Jacobson:2010mx}. By doing so, the corresponding SME-based Lagrange density is readily obtained as
\begin{subequations}
\begin{equation}
\mathcal{L}^{(2)}_{\mathrm{gravity}}=\frac{1}{2\kappa}\breve{k}^{(2)}\,,\quad \breve{k}^{(2)}=k_{\kappa\lambda\mu\nu}g^{\kappa\mu}g^{\lambda\nu}+k_{\mu\nu}g^{\mu\nu}\,,
\end{equation}
with the background fields
\begin{align}
k_{\kappa\lambda\mu\nu}&=-\left(1+\frac{\lambda}{\xi}\right)\nabla_{\kappa}u_{\mu}\nabla_{\lambda}u_{\nu}+\left(1+\frac{1}{\xi}\right)\nabla_{\kappa}u_{\lambda}\nabla_{\nu}u_{\mu}\,, \\[2ex]
k_{\mu\nu}&=-\frac{\alpha}{\xi}u^{\kappa}u^{\lambda}\nabla_{\kappa}u_{\mu}\nabla_{\lambda}u_{\nu}\,,
\end{align}
\end{subequations}
where $\xi=G_H/G_N$ and $u^{\mu}$ is a unit vector orthogonal to the spacelike hypersurfaces of the spacetime foliation. Here we refer to the second line of Table~XVI and eq.~(41) of ref.~\cite{Kostelecky:2021xhb}. Note that the author of ref.~\cite{Jacobson:2010mx} uses the metric signature $(+,-,-,-)$, which is different from that of ref.~\cite{Kostelecky:2021xhb}.

%................................................................................
\section{Conclusions}
\label{sec:conclusions}
%................................................................................
This work was dedicated to a modified-gravity theory incorporating nondynamical background fields that break diffeomorphism invariance explicitly. We focused on two distinct types of background fields: a scalar field denoted as $u$ and a tensor-valued one known as $s^{\mu\nu}$ in the SME literature \cite{Kostelecky:2003fs,Kostelecky:2020hbb}. The latter decomposes into a purely timelike and a purely spacelike sector. The canonical Hamiltonian had already been obtained in the previous article~\cite{Reyes:2021cpx} and formed the base of the current paper.

On the one hand, the ADM decomposition allowed us to project the modified Einstein equations into purely spacelike hypersurfaces. On the other hand, the Hamilton-Jacobi equations gave rise to another set of field equations for the theory. We have found that both approaches are consistent with each other. This outcome is highly nontrivial, as nondynamical background fields transform unconventionally under diffeomorphisms. Hence, the structure of such theories is rendered substantially obscure as opposed to GR or extensions thereof with either diffeomorphism symmetry intact or broken spontaneously. Such an analysis in the context of the gravitational SME is the first of its kind in the literature. In principle, the Dirac procedure could imply additional sets of constraints beyond those of GR. However, our findings indicate that the canonical Hamiltonian completely encodes the dynamical properties of the gravitational theory under study. Furthermore, they corroborate the results derived previously in ref.~\cite{Reyes:2021cpx}.

The setting investigated is quite generic and links to several popular modified-gravity models proposed in the literature such as Brans-Dicke theory and dRGT massive gravity. Hence, this investigation could also arouse interest of scientists not having had contact to the gravitational SME, so far. Additional explorations of the constraint structure are beyond the scope of this paper and will be pursued in a future work. Gaining a better understanding of the SME sectors identified in sections~\ref{eq:linearized-dRGT-theory}, \ref{sec:horava-lifshitz-gravity} by utilizing the methodology employed here may be further worthwhile projects.

\section{Acknowledgments}

It is a pleasure to thank P.~Sundell for valuable discussions on several aspects of this project. C.M.R acknowledges partial support by the research project Fondecyt Regular 1191553. M.S. is indebted to FAPEMA Universal 00830/19, CNPq Produtividade 312201/2018-4, and CAPES/Finance Code 001.

%..........................................................................
\bibliographystyle{JHEP}

%\bibitem{a}
%Author, \emph{Title}, \emph{J. Abbrev.} {\bf vol} (year) pg.
%
%\bibitem{b}
%Author, \emph{Title},
%arxiv:1234.5678.
%
%\bibitem{c}
%Author, \emph{Title},
%Publisher (year).
%
%
%% Please avoid comments such as "For a review'', "For some examples",
%% "and references therein" or move them in the text. In general,
%% please leave only references in the bibliography and move all
%% accessory text in footnotes.
%
%% Also, please have only one work for each \bibitem.
%
%

\end{document}